\documentclass[a4paper,preprintnumbers,showpacs,twocolumn,superscriptaddress,nofootinbib,amsmath,amssymb]{revtex4-1}
\usepackage[dvips]{graphics}
\usepackage{times}
\usepackage{mathrsfs}
\usepackage[hypertex]{hyperref}
\usepackage{color,hyperref}
\hypersetup{
  colorlinks=true,        % false: boxed links; true: colored links
  linkcolor=blue,         % color of internal links
  citecolor=magenta,      % color of links to bibliography
  filecolor=magenta,      % color of file links
  urlcolor=blue            % color of external links
}

\def\beq{\begin{equation}}
\def\eeq{\end{equation}}
\def\bear{\begin{eqnarray}}
\def\ear{\end{eqnarray}}

\def\L{\mathscr{L}}

\usepackage{enumitem}
\usepackage{graphicx,subfigure}% Include figure files
\usepackage{dcolumn}% Align table columns on decimal point
\usepackage{bm}% bold math
\usepackage{color}

\begin{document}

\title{Electromagnetic perturbations of black holes in general relativity coupled to nonlinear electrodynamics: Polar perturbations}

\author{Bobir Toshmatov}
\email{bobir.toshmatov@fpf.slu.cz}

\affiliation{Institute of Physics and Research Centre of Theoretical  Physics and Astrophysics, Faculty of Philosophy \& Science, Silesian University in Opava, Bezru\v{c}ovo n\'{a}m\v{e}st\'{i} 13, CZ-74601 Opava, Czech Republic}

\affiliation{Department of Physics, Nazarbayev University, 53 Kabanbay Batyr, 010000 Astana, Kazakhstan}

\affiliation{Ulugh Beg Astronomical Institute, Astronomicheskaya 33, Tashkent 100052, Uzbekistan}

\author{Zden\v{e}k Stuchl\'{i}k}
\email{zdenek.stuchlik@fpf.slu.cz}
\affiliation{Institute of Physics and Research Centre of  Theoretical Physics and Astrophysics, Faculty of Philosophy \& Science, Silesian University in Opava, Bezru\v{c}ovo n\'{a}m\v{e}st\'{i} 13, CZ-74601 Opava, Czech Republic}

\author{Bobomurat Ahmedov}
\email{ahmedov@astrin.uz}
\affiliation{Ulugh Beg Astronomical Institute, Astronomicheskaya 33, Tashkent 100052, Uzbekistan}

\affiliation{National University of Uzbekistan, Tashkent 100174,
Uzbekistan}

\begin{abstract}

The \textit{axial} electromagnetic (EM) perturbations of the black hole (BH) solutions in general relativity coupled to nonlinear electrodynamics (NED) were studied for both electrically and magnetically charged BHs, assuming that the EM
perturbations do not alter the spacetime geometry in our preceding paper [Phys. Rev. D 97, 084058 (2018)]. Here, as a continuation of that work, the formalism for the \textit{polar} EM perturbations of the BHs in general relativity coupled to the NED is presented. We show that the quasinormal modes (QNMs) spectra of polar EM perturbations of the electrically and magnetically charged BHs in the NED are not isospectral, contrary to the case of the standard Reissner-Nordstr\"{o}m BHs in the classical linear electrodynamics. It is shown by the detailed study of QNMs properties in the eikonal approximation that the EM perturbations can be a powerful tool to confirm that in the NED light ray does not follow the null geodesics of the spacetime. By specifying the NED model and comparing axial and polar EM perturbations of the electrically and magnetically charged BHs, it is shown that QNM spectra of the axial EM perturbations of magnetically (electrically) charged BH and polar EM perturbations of the electrically (magnetically) charged BH are isospectral, i.e., $\omega_{mag}^{ax}\approx\omega_{el}^{pol}$ ($\omega_{mag}^{pol}\approx\omega_{el}^{ax}$).

\end{abstract}

\maketitle

\section{Introduction}\label{sec-intr}

Black holes (BHs) are among the simplest and at the same time most bizarre objects in the Universe -- they have only three defining attributes, mass, spin, and electric (or magnetic) charge, according to the \textit{no-hair theorem}~\cite{MTW1973}, and there is a spacetime singularity in their interior that is enclosed by an event horizon according to the \textit{cosmic censorship}~\cite{Penrose1969}. By listing those three parameters accordingly, one can depict a complete portrait of the BH environment. However, we do not yet have a good enough theory of gravity to explain and describe the spacetime singularity. Therefore, special interests have been raised in coupling general relativity (GR) to another fundamental field theories, such as nonlinear electrodynamics (NED)~\footnote{Coupling general relativity to the linear or Maxwell electrodynamic field gives the Reissner-Nordstr\"{o}m (RN) black hole solution which is singular at the origin of the spacetime $r=0$.}, to obtain the singularity-free BH solutions~\cite{ABG:PRL:1998,ABG:PLB:2000,Bronnikov:PRD:2001,Dymnikova:CQG:2004,Dymnikova:CQG2015,Rodrigues:PRD2018}. The generic class of singular and singularity-free BH solutions in GR coupled to the NED is presented in Ref.~\cite{Fan:PRD:2016} and refined in Refs.~\cite{Bronnikov:PRD2017,TSA:PRDnew}. Such singularity-free BHs are called \textit{regular BHs}.

Although BHs cannot be seen directly, one can guess their presence in the particular place of the space by measuring their strong gravity effects on the surrounding objects: mass estimates from test objects orbiting or spiraling into a BH, gravitational lens effects, and radiation emitted by the surrounding matter. Apart from these effects, we can ``hear" their collisions representing the final stage of the evolution of the close BH binaries. As sound waves disturb the air to make noise, gravitational waves (GWs) disturb the fabric of spacetime to push and pull matter, as recently LIGO and VIRGO global experiments have directly detected the GWs from the coalescence of two BHs~\cite{LIGOPRL:2016}. The coalescence of two BHs can be divided into three stages: the inspiral, merger and ringdown. Each phase can be calculated by different means. The inspiral can be studied analytically within the post-Newtonian approximation, while the merger is directly computable by using the numerical relativity only. Finally, the ringdown phase describes relaxation of the final object to an equilibrium state by emitting GWs in so-called quasinormal modes (QNMs), the frequencies of this are complex, giving thus also damping of the oscillations. This phase can be also calculated analytically via perturbation theory (see Refs.~\cite{Kokkotas,Rezzolla:review,Berti:review,Konoplya:review} and references therein).

In the recent preceding paper~\cite{TSSA:PRD:2018}, we studied the behavior of the dynamical response of the spherically symmetric, magnetically and/or electrically charged BHs representing exact solutions of coupled Einstein's gravity and the NED to the \emph{axial} electromagnetic (EM) perturbations, assuming the EM perturbations do not alter the spacetime geometry. One of the main reasons for the topic was to determine whether it is possible to distinguish the BHs related to the NED from the BHs related to the standard linear electrodynamics (LED) from their response to the EM perturbations. In that paper we showed that i) the axial EM perturbations of in the NED BHs give different potentials and, consequently, different QNM spectra, in comparison with those related to the RN BHs governed by the standard LED, since it is well known from Refs.~\cite{ChaverraPRD:93,SarbachPRD:67,TSSA:PRD:2018} that the QNMs of the EM perturbations of the electrically and magnetically charged RN BHs are isospectral, with identical effective potentials; ii) in the eikonal (large multipole numbers) regime, the QNMs of the NED BHs are determined by the unstable circular photon orbits determined by the given geometry, i.e., by unstable circular null geodesics determined by the effective (or optical) geometry, since in the NED light ray does not follow the null geodesics of the spacetime~\cite{Novello:PRD:2000,Novello:PRD:2001,ObukhovPRD:66,Breto:PRD:2016,deOlivieraCQG:26,StuchlikIJMPD:24,Schee:JCAP2015,Schee:CQG2016}.
It should be noted that one of the outstanding pioneering works devoted to the study of the perturbations of the one-parameter family of Lagrangian densities that yields static, spherically symmetric magnetically charged BH solutions in the NED is the paper~\cite{ChaverraPRD:93} that overlaps some important results of the present paper and Ref.~\cite{TSSA:PRD:2018}, despite the different models that were used. In that paper the following very important results have been presented:
i) A stability analysis has been presented.
ii) By studying the QNM spectra it has been shown that the even-parity and odd-parity perturbations are not isospectral in the NED.
iii) For the eikonal limit it has been shown that the unstable circular photon orbits of the spacetime plays an important role.

In this paper, as a continuation of our preceding paper~\cite{TSSA:PRD:2018}, we study within this framework the \emph{polar} EM perturbations of the spherically symmetric, magnetically and electrically charged BHs representing exact solutions of coupled Einstein's gravity (GR) and NED. The paper is organized as follows. In Sec.~\ref{sec-intro} we review the equations of motion governing a self-gravitating, NED configuration and discuss spherically symmetric, magnetically and electrically charged BH solutions. In Sec.~\ref{sec-perturb} we demonstrate the \emph{polar} EM perturbations of the electrically and magnetically charged, spherically symmetric NED BHs. In Sec.~\ref{sec-eikonal} we study the QNMs of the electrically charged BHs in the large multipole numbers limit. We apply the obtained formalism for the specific type of BHs in GR coupled to the NED, calculate their QNMs in comparison with the ones of the Schwarzschild and RN BHs in Sec.~\ref{sec-application}. Finally, in Sec.~\ref{sec-conclude} we summarize the main results. In this paper we mainly use the  natural units $\hbar=c=G=1$. Furthermore, we adopt $(-, +, +, +)$ convention for the signature of the metric.

\section{BH solutions in GR coupled to NED}\label{sec-intro}

In general in the case of GR coupled to NED, the action is given by
\begin{eqnarray}\label{action}
S = \frac{1}{16 \pi }\int d^4 x \sqrt{ -g }\left( R - \mathscr{L} \right)\ ,
\end{eqnarray}
where $g$ is the determinant of the metric tensor, $R$ is the Ricci scalar, and $\L$ is the Lagrangian density describing the NED theory that depends on $\L=\L(F\equiv F_{\mu\nu}F^{\mu\nu})$, with $F_{\mu\nu}=\partial_\mu A_\nu-\partial_\nu A_\mu$. Since $F_{\mu\nu}$ is antisymmetric, it has only six
nonzero components.

The covariant equations of motion are written in the form
\begin{eqnarray}
&&G_{\mu\nu}=T_{\mu\nu},  \label{eq-motion1}\\
&&\nabla_\nu\left(\L_F F^{\mu\nu}\right)=0\ ,   \label{eq-motion2}
\end{eqnarray}
where $G_{\mu\nu}=R_{\mu\nu}-Rg_{\mu\nu}/2$  and $T_{\mu\nu}$ are the Einstein tensor and the energy-momentum tensor of the NED field, respectively. The energy-momentum tensor of the NED is determined by the relation
\begin{eqnarray}\label{em-tensor}
T_{\mu \nu} = 2 \left( \L_F
F_\mu^\alpha F_{\nu\alpha}
- \frac{1}{4}
g_{\mu \nu} \L \right)\ ,
\end{eqnarray}
where $\L_F=\partial_F\L$.

Let us consider the line element of the static, spherically symmetric BH is given in the form
\begin{eqnarray}
ds^2 = - f( r) d t^2 + \frac{ dr^2}{f (r)}
+r^2 \left( d \theta^2+ \sin^2 \theta  d\phi^2 \right)\ , \label{line element}
\end{eqnarray}
where GR and NED evaluate the lapse function $f(r)$. The line element~(\ref{line element}) satisfies the symmetry $G_t^t=G_r^r$.

In general, the EM 4-potential can be written in the following form:
\begin{eqnarray}\label{ansatz}
\bar{A}_\mu=\varphi(r)\delta_\mu^t-Q_m\cos\theta\delta_\mu^\phi\ ,
\end{eqnarray}
where $\varphi(r)$ and $Q_m$ are the electric potential and the total magnetic charge, respectively.
Below based on the
method of Bronnikov~\cite{Bronnikov:PRD:2001} we briefly demonstrate the
formalism of construction of electrically
and magnetically charged BHs in
GR coupled to the NED.

Since the formalism of construction of the electrically and magnetically charged BHs
in GR coupled to the NED has already been presented in~\cite{Bronnikov:PRD:2001,Fan:PRD:2016,TSSA:PRD:2018}, we do not report
the whole procedure in detail,
instead, we briefly mention some key moments.

The electrically charged BH solution with the ansatz $\bar{A}_t=\varphi(r)$, and the
EM field strength $F=-2\varphi'^2$,
can be constructed by solving the Einstein field equations~(\ref{eq-motion1}) as
\bear
&&\L = \frac{2 m''}{ r}, \label{Lagrangian-el1}\\
&&\L_F = \frac{2 m' - r m''}{2 r^2 \varphi'^2}\ ,\label{Lagrangian-el2}
\ear
where $m$ is the radially and
EM field dependent mass function of the Schwarzschild-like BHs, related to the metric function as
\bear\label{metric-function}
f (r) = 1 - \frac{2 m (r)}{ r}\ .
\ear
From Eqs.~(\ref{Lagrangian-el1})
and~(\ref{Lagrangian-el2}) one can easily
notice that if $m(r)=M$, the Lagrangian
density of the NED vanishes, $\L=0$, and one arrives at the Schwarzschild solution
of GR. The total electric charge
inside the sphere with radius $r$
is found by equations of motion~(\ref{eq-motion2}) as
\bear\label{total-charge}
Q_e
=
r^2 \L_F \varphi'\ .
\ear
By substituting~(\ref{Lagrangian-el2}) %
to~(\ref{total-charge}), and solving %
the differential equation with
respect to the electric
potential $\varphi(r)$, one
obtains the expression for the electric potential as
\bear\label{potential-el}
\varphi = \frac{3m - rm'}{2 Q_e} + C\ ,
\ear
where $C$ is an integration constant.
If one considers the EM field is
linear, $\L=F$ (or $\L_F=1$), the differential
equation (\ref{Lagrangian-el1})
(or (\ref{Lagrangian-el2})) gives the RN
BH spacetime with mass function
$m(r)=M+Q_e^2/2r$, and corresponding electric
potential~(\ref{potential-el}) takes
the form $\varphi=Q_e/r$.

The magnetically charged spherically symmetric BH solution with the ansatz $\bar{A}_\phi=-Q_m\cos\theta$ and the EM field strength $F=2Q_m^2/r^4$ can be constructed by solving the Einstein field equations~(\ref{eq-motion1}) as
\bear
&&\L = \frac{ 4 m'}{ r^2}\ , \label{Lagrangian-mg1}\\
&&\L_F = \frac{ r^2 (2 m' - r m'')}{2 Q_m^2}\ . \label{Lagrangian-mg2}
\ear
If one considers the EM field is linear, $\L=F$ (or $\L_F=1$), the differential equation (\ref{Lagrangian-mg1}) (or (\ref{Lagrangian-mg2})) gives the RN BH spacetime with mass function $m(r)=M+Q_e^2/2r$.

By choosing the mass function related to the electric or magnetic fields as presented above, one can construct BH solutions in GR coupled to the NED. One more important property of the NED is that the NED can eliminate the curvature singularity (divergence of the curvature) of the spacetime. For details, see Refs.~\cite{Fan:PRD:2016,TSSA:PRD:2018}.

\section{Polar EM perturbations of BHs in GR coupled to the NED}\label{sec-perturb}

In this section we study polar EM perturbations of BHs in
NED by introducing the polar EM perturbations into gauge
potential~(\ref{ansatz}) as
\begin{eqnarray}
A_\mu = \bar{A }_\mu +
\delta A_\mu\ ,\label{ansatz-perturb}
\end{eqnarray}
considering the polar perturbations given in the form
\begin{eqnarray}\label{axial-perturb}
\delta A_\mu&=&\sum_{\ell,m}\left(\left[
\begin{array}{c}
d^{\ell m}(t,r)Y_{\ell m}(\theta,\phi)\\
h^{\ell m}(t,r)Y_{\ell m}(\theta,\phi)\\
k^{\ell m}(t,r)\partial_\theta Y_{\ell m}(\theta,\phi)\\
k^{\ell m}(t,r)\partial_\phi Y_{\ell m}(\theta,\phi)\\
\end{array}\right]
\right)\ ,
\end{eqnarray}
where $Y_{\ell m}(\theta,\phi)$
is the spherical harmonic function
of degree $\ell$ and
order $m$ related to the
angular coordinates $\theta$ and
$\phi$. In the next subsection
we study electrically and
magnetically charged
BH cases separately.

\subsection{Magnetically charged black holes}

The EM four-potential of the magnetically charged BH is given as $\bar{A}_\mu=-Q_m\cos\theta\delta_\mu^\phi$. Then, nonzero components of the EM field tensor are given as
\bear\label{cov-magnetic}
F_{tr}&&=\left(\partial_th^{\ell m}-\partial_rd^{\ell m}\right)Y_{\ell m}\ , \nonumber\\
F_{t\theta}&&=\left(\partial_tk^{\ell m}-d^{\ell m}\right)\partial_\theta Y_{\ell m}\ , \nonumber\\
F_{t\phi}&&=\left(\partial_tk^{\ell m}-d^{\ell m}\right)\partial_\phi Y_{\ell m}\ , \nonumber\\
F_{r\theta}&&=(\partial_r k^{\ell m}-h^{\ell m})\partial_\theta Y_{\ell m}\ ,\\
F_{r\phi}&&=(\partial_r k^{\ell m}-h^{\ell m})\partial_\phi Y_{\ell m}\ ,\nonumber\\
F_{\theta\phi}&&=Q_m\sin\theta\ .\nonumber
\ear

The contravariant nonzero components of the EM field tensor are written using the relation
$F^{\mu\nu}=g^{\mu\alpha}g^{\nu\beta}F_{\alpha\beta}$ as
\bear\label{contr-magnetic}
F_{tr}&&=-\left(\partial_th^{\ell m}-\partial_rd^{\ell m}\right)Y_{\ell m}\ , \nonumber\\
F_{t\theta}&&=-\frac{1}{r^2f}\left(\partial_tk^{\ell m}-d^{\ell m}\right)\partial_\theta Y_{\ell m}\ , \nonumber\\
F_{t\phi}&&=-\frac{1}{r^2f\sin^2\theta}\left(\partial_tk^{\ell m}-d^{\ell m}\right)\partial_\phi Y_{\ell m}\ , \nonumber\\
F_{r\theta}&&=\frac{f}{r^2}(\partial_r k^{\ell m}-h^{\ell m})\partial_\theta Y_{\ell m}\ ,\\
F_{r\phi}&&=\frac{f}{r^2\sin^2\theta}(\partial_r k^{\ell m}-h^{\ell m})\partial_\phi Y_{\ell m}\ ,\nonumber\\
F_{\theta\phi}&&=\frac{Q_m}{r^4\sin\theta}\ .\nonumber
\ear
In the linear perturbations (up to the first order) approximation the EM field strength remains unchanged as $F=\bar{F}$, where
\bear\label{EM-strength}
\bar{F}=\frac{2Q_m^2}{r^4}\ ,
\ear
Consequently, the lagrangian density of the NED also remains unchanged in the linear perturbations approximation, $\L_F=\bar{\L}_{\bar{F}}$.
With the above given expressions one can obtain from~(\ref{eq-motion2}) the following three independent ordinary differential equations (ODEs):
\bear
&&rf\left(2+\frac{r\L_F'}{\L_F}\right)(\partial_th^{\ell m}-\partial_rd^{\ell m})\label{t-eq}\\&&+ r^2f\partial_r(\partial_th^{\ell m}-\partial_rd^{\ell m})-\ell(\ell+1)(\partial_t k^{\ell m}-d^{\ell m})=0\ ,\nonumber\\
&&\frac{r^2}{f}(\partial_th^{\ell m}-\partial_rd^{\ell m})-\ell(\ell+1)(\partial_t k^{\ell m}-h^{\ell m})=0\ ,\label{r-eq}\\
&&f(f\L_F)'(\partial_rk^{\ell m}-h^{\ell m})+f^2\L_F\partial_r(\partial_rk^{\ell m}-h^{\ell m})\nonumber\\&&+\L_F\partial_t(d^{\ell m}-\partial_tk^{\ell m})=0\ ,\label{phi-eq}
\ear
where Eqs.~(\ref{t-eq}),~(\ref{r-eq}),~(\ref{phi-eq}) correspond to $\mu=t$, $\mu=r$, $\mu=\theta=\phi$, respectively. By differentiating Eqs.~(\ref{t-eq}) and~(\ref{r-eq}) with respect to $r$ and $t$, respectively, and introducing the tortoise (Regge-Wheeler) coordinate, $dx=dr/f$, and new variable
\bear
\Psi^{\ell m}(t,r)=\frac{1}{r^2\sqrt{\L_F}}(\partial_th^{\ell m}-\partial_rd^{\ell m})\ ,
\ear
one arrives at the well-known wave equation~\footnote{In~\cite{Molina:PRD:2016,ToshmatovPRD:96} alternate method of derivation of the wave equation~(\ref{wave-eq1}) from (\ref{phi-eq}) is presented.}
\bear\label{wave-eq1}
\left[-\frac{\partial^2}{\partial t^2}+\frac{\partial^2}{\partial
x^2}-V_{mag}^{pol}(r)\right]\Psi(t,r)=0\ ,
\ear
where the effective potential is defined by the expression
\bear\label{pot-mag1}
V_{mag}^{pol}(r)=f\left[\frac{\ell(\ell+1)}{r^2}-\frac{3f\L_F'^2-2\L_F\left(f\L_F'\right)'}{4\L_F^2}\right]\ .
\ear
If the EM field is linear, $\L_F=1$, then, one recovers the well-known effective potential $V_{mag}^{pol}=f\ell(\ell+1)/r^2$ which corresponds to the standard RN and other BHs which are not related to the electrodynamics ($\L_F=0$)~\cite{TSSA:PRD2016,TASA:PRD2015}.

\subsection{Electrically charged black holes}

The EM four-potential of the electrically charged BH is given as $\bar{A}_\mu=\varphi(r)\delta_\mu^t$. Then, nonzero components of the EM field tensor are given as
\bear\label{cov-electric}
F_{tr}&&=-\varphi'+\left(\partial_th^{\ell m}-\partial_rd^{\ell m}\right)Y_{\ell m}\ , \nonumber\\
F_{t\theta}&&=\left(\partial_tk^{\ell m}-d^{\ell m}\right)\partial_\theta Y_{\ell m}\ , \nonumber\\
F_{t\phi}&&=\left(\partial_tk^{\ell m}-d^{\ell m}\right)\partial_\phi Y_{\ell m}\ , \\
F_{r\theta}&&=(\partial_r k^{\ell m}-h^{\ell m})\partial_\theta Y_{\ell m}\ ,\nonumber\\
F_{r\phi}&&=(\partial_r k^{\ell m}-h^{\ell m})\partial_\phi Y_{\ell m}\ ,\nonumber
\ear
The contravariant nonzero components of the EM field tensor are written using the relation
$F^{\mu\nu}=g^{\mu\alpha}g^{\nu\beta}F_{\alpha\beta}$ as
\bear\label{contra-electric}
F^{tr}&&=\varphi'-\left(\partial_th^{\ell m}-\partial_rd^{\ell m}\right)Y_{\ell m}\ , \nonumber\\
F^{t\theta}&&=-\frac{1}{r^2f}\left(\partial_tk^{\ell m}-d^{\ell m}\right)\partial_\theta Y_{\ell m}\ , \nonumber\\
F^{t\phi}&&=-\frac{1}{r^2f\sin^2\theta}\left(\partial_tk^{\ell m}-d^{\ell m}\right)\partial_\phi Y_{\ell m}\ , \\
F^{r\theta}&&=\frac{f}{r^2}(\partial_r k^{\ell m}-h^{\ell m})\partial_\theta Y_{\ell m}\ ,\nonumber\\
F^{r\phi}&&=\frac{f}{r^2\sin^2\theta}(\partial_r k^{\ell m}-h^{\ell m})\partial_\phi Y_{\ell m}\ ,\nonumber
\ear
Thus the EM field strength also takes new form as $F=\bar{F}+\delta F$ where
\bear\label{EM-strength-electric}
\bar{F}=-2\varphi'^2,\quad \delta F=4\varphi'\left(\partial_th^{\ell m}-\partial_rd^{\ell m}\right)Y_{\ell m}\ .
\ear
Consequently, $\L_F$ also changes to $\L_F=\bar{\L}_{\bar{F}}+\bar{\L}_{\bar{F}\bar{F}}\delta F$. By using this and the contravariant components of the EM field tensor~(\ref{contra-electric}) in Eq.~(\ref{eq-motion2}), one arrives at the following ODEs:
\begin{widetext}
\bear
&&\frac{rf[2\bar{\L}_{\bar{F}}+r\bar{\L}_{\bar{F}}'- 4\varphi'\left(r\bar{\L}_{\bar{F}}'\varphi'+2\bar{\L}_{\bar{F}\bar{F}}(r\varphi')' \right)]}{\bar{\L}_{\bar{F}}}\left(\partial_th^{\ell m}-\partial_rd^{\ell m}\right)\nonumber\\&&+r^2f(\bar{\L}_{\bar{F}}-4\bar{\L}_{\bar{F}\bar{F}}\varphi'^2)\partial_r \left(\partial_th^{\ell m}-\partial_rd^{\ell m}\right)-\ell(\ell+1)(\partial_tk^{\ell m}-d^{\ell m})=0\ ,\label{t-eq-pol}\\
&&\frac{r^2(\bar{\L}_{\bar{F}}-4\bar{\L}_{\bar{F}\bar{F}}\varphi'^2)}{f\bar{\L}_{\bar{F}}}\partial_t \left(\partial_th^{\ell m}-\partial_rd^{\ell m}\right)-\ell(\ell+1)(\partial_rk^{\ell m}-h^{\ell m})=0\ ,\label{r-eq-pol}
%&&\bar{\L}_{\bar{F}}\left[(\partial_td^{\ell m}-\partial_t^2k^{\ell m})-f^2(\partial_rd^{\ell m}-\partial_r^2k^{\ell m})\right]+f(f\bar{\L}_{\bar{F}})'(\partial_rk^{\ell m}-h^{\ell m})=0\ ,\label{th-eq-pol}
\ear
\end{widetext}
Differentiating Eqs.~(\ref{t-eq-pol}) and~(\ref{r-eq-pol}) with respect to the coordinates $r$ and $t$, respectively, and subtracting them, one arrives at the well known wave equation
\bear\label{wave-eq1}
\left[-\frac{\partial^2}{\partial t^2}+\frac{\partial^2}{\partial
z^2}-V_{el}^{pol}(r)\right]\Phi(t,r)=0\ ,
\ear
where the new function $\Phi(t,r)$ is introduced  and defined by
\bear
\Phi^{\ell m}(t,r)=\partial_th^{\ell m}-\partial_rd^{\ell m}\ ,
\ear
and $z$ is newly introduced tortoise-like coordinate, $dz=dr/f\sqrt{\bar{\L}_{\bar{F}}}$. The potential $V_{el}^{pol}$ is given by
\begin{widetext}
\bear\label{potential-el1}
&&V_{el}^{pol}(r)=\frac{f}{\left(\bar{\L}_{\bar{F}}-4\bar{\L}_{\bar{F}\bar{F}}\varphi'^2\right)} \left[\frac{\bar{\L}_{\bar{F}}\ell(\ell+1)}{r^2} -\frac{f'A}{4r}-\frac{fB}{16r^2\left(\bar{\L}_{\bar{F}}-4\bar{\L}_{\bar{F}\bar{F}}\varphi'^2\right) \bar{\L}_{\bar{F}}}\right]\ ,
\ear
\end{widetext}
where
\begin{widetext}
\bear
A&&=-8(\bar{\L}_{\bar{F}}-1)\varphi'\left[2\bar{\L}_{\bar{F}\bar{F}}(r\varphi')'+ \bar{\L}_{\bar{F}\bar{F}}'r\varphi' \right]+4\bar{\L}_{\bar{F}}^2+\bar{\L}_{\bar{F}}(r\bar{\L}_{\bar{F}}'-4) +2r\bar{\L}_{\bar{F}}'(2\bar{\L}_{\bar{F}\bar{F}}\varphi'^2-1)\ ,\nonumber\\
B&&=-4\left[r \bar{\L}_{\bar{F}}'\left(2 \bar{\L}_{\bar{F}\bar{F}}\varphi'^2-1\right)+4\varphi'\left(r\bar{\L}_{\bar{F}\bar{F}}'\varphi'+2 \bar{\L}_{\bar{F}\bar{F}}(r\varphi')'\right)\right]\left[r\bar{\L}_{\bar{F}}' \left(6 \bar{\L}_{\bar{F}\bar{F}}\varphi'^2-1\right)+4\varphi'\left(r \bar{\L}_{\bar{F}\bar{F}}' \varphi'+2\bar{\L}_{\bar{F}\bar{F}}(r\varphi')'\right)\right]\nonumber\\ &&+8 \bar{\L}_{\bar{F}}\left\{4\varphi'\left[\bar{\L}_{\bar{F}\bar{F}}\left(\varphi'\left(r\left(4 \varphi'^2 \left(r \bar{\L}_{\bar{F}\bar{F}}''+4\bar{\L}_{\bar{F}\bar{F}}'\right)-r \bar{\L}_{\bar{F}}''\right)+4\right)+4 r\varphi''\left(4 r \bar{\L}_{\bar{F}\bar{F}}' \varphi'^2+1\right)\right)+2 \bar{\L}_{\bar{F}\bar{F}}^2\varphi'\left(\left(r^2 \bar{\L}_{\bar{F}}''+4\right)\varphi'^2\right.\right.\right.\nonumber\\&&\left.\left.\left.+4r^2\varphi''^2+4 r\varphi' \left(r \varphi^{(3)}+4\varphi''\right)\right)+2 r\bar{\L}_{\bar{F}\bar{F}}'\varphi'\right]+2 r \bar{\L}_{\bar{F}}'\left(2r\varphi'\left(\bar{\L}_{\bar{F}\bar{F}}'\varphi'+2\bar{\L}_{\bar{F}\bar{F}}\varphi''\right)-1\right)+r^2 \bar{\L}_{\bar{F}}'^2 \left(3\bar{\L}_{\bar{F}\bar{F}}\varphi'^2-1\right)\right\}\nonumber\\&&+4
\bar{\L}_{\bar{F}}^3 \left\{r\left[-r \bar{\L}_{\bar{F}}''-4 \bar{\L}_{\bar{F}}'+8\left(r \bar{\L}_{\bar{F}\bar{F}}''\varphi'^2+2\bar{\L}_{\bar{F}\bar{F}}'\varphi'\left(2r\varphi''+\varphi'\right)+2 r\bar{\L}_{\bar{F}\bar{F}}\varphi''^2+2\bar{\L}_{\bar{F}\bar{F}}\varphi'\left(r\varphi^{(3)} +2\varphi''\right)\right)\right]+4\right\}\nonumber\\&&+\bar{\L}_{\bar{F}}^2
   \left\{r \left[-8 \left(-r\bar{\L}_{\bar{F}}''+4 \left(4 r \bar{\L}_{\bar{F}\bar{F}}\bar{\L}_{\bar{F}\bar{F}}''\varphi'^4+r\bar{\L}_{\bar{F}\bar{F}}''\varphi'^2+2r \bar{\L}_{\bar{F}\bar{F}}\varphi''^2+8\bar{\L}_{\bar{F}\bar{F}}^2\varphi'^3 \left(r\varphi^{(3)}+2\varphi''\right)\right.\right.\right.\right.\nonumber\\&&\left.\left.\left.\left.+2 \bar{\L}_{\bar{F}\bar{F}}\varphi'\left(r\varphi^{(3)}+4\varphi''\right)\right)-8 r\bar{\L}_{\bar{F}\bar{F}}'^2\varphi'^4+16 \bar{\L}_{\bar{F}\bar{F}}'\varphi'\left(2 \bar{\L}_{\bar{F}\bar{F}}\varphi'^2+1\right) \left(r\varphi''+\varphi'\right)\right)\right.\right.\nonumber\\&&\left.\left.+16 \bar{\L}_{\bar{F}}'\left(-2r\bar{\L}_{\bar{F}\bar{F}}'\varphi'^2+4 \bar{\L}_{\bar{F}\bar{F}}\varphi' \left(\varphi'-r\varphi''\right)+1\right)+r
 \bar{\L}_{\bar{F}}'^2\right]-16 \left(8 \bar{\L}_{\bar{F}\bar{F}}\varphi'^2+1\right)\right\}\ .\nonumber
%A=\bar{\L}_{\bar{F}}\left[r \bar{\L}_{\bar{F}}'-8\varphi'\left(2\bar{\L}_{\bar{F}\bar{F}}(r\varphi')'+ \bar{\L}_{\bar{F}\bar{F}}'r\varphi' \right)-4\right]+2 r\bar{\L}_{\bar{F}}'\left(2\bar{\L}_{\bar{F}\bar{F}}\varphi'^2-1\right)+4\bar{\L}_{\bar{F}}^2+8\varphi' \left(2\bar{\L}_{\bar{F}\bar{F}}(r\varphi')'+ \bar{\L}_{\bar{F}\bar{F}}'r\varphi' \right),\nonumber
\ear
\end{widetext}
In the case of the linear EM field, $\bar{\L}_{\bar{F}}=1$, the effective potential (\ref{potential-el1}) turns out the one for the standard RN BHs.

Since solving the wave equation (\ref{wave-eq1}) with the potential (\ref{potential-el1}) requires to specify the NED (lagrangian density), we postpone these calculations to the next sections where we plan to present few special models.

\section{QNMs of NED BHs from unstable null geodesics}\label{sec-eikonal}

According to~\cite{CardosoPRD:79}, the QNMs of any stationary, spherically symmetric and asymptotically flat BHs in any dimensions are determined in the eikonal regime~\footnote{Here, eikonal regime means the large multipole numbers limit.} by the circular null geodesics~\footnote{However, in~\cite{KonoplyaPLB:2017,KonoplyaJCAP:2017} it has been shown for the Einstein-Lovelock theory that the relation (\ref{eikonal1}) is not universal feature of all stationary, spherically symmetric and asymptotically flat BHs in any dimensions.} namely, the real part of the QN frequencies is determined by angular velocity of the unstable null geodesics, $\Omega_c$, while the imaginary part of the QN frequencies is determined by the so-called Lyapunov exponent, $\lambda$ as
\bear
\omega = \Omega_c\ell - i\left( n + \frac{1}{2} \right) | \lambda |\ ,\label{eikonal1}
\ear
where $\Omega_c$ and $\lambda$ for the spacetime metric~(\ref{line element}) are given by the following expressions
\bear
&&\Omega_c = \sqrt{\frac{f_c}{r_c^2}}\ ,\label{omega_c}\\
&&\lambda = \sqrt{-\frac{r_c^2 }{ 2 f_c}\left(\frac{d ^2}{dx ^2}
\frac{f}{r^2}\right)|_{ r = r_c }}\ ,\label{lambda}
\ear
where $x$ is
the tortoise coordinate,
$r_c$ is radius of the
unstable null circular orbit
which is determined by the equation
$r_cf_c'-2f_c=0$.
However, in our preceding paper~\cite{TSSA:PRD:2018}, we have shown in the study of the \textit{axial} EM perturbations to the BHs in GR coupled to the NED that the QNMs of any stationary, spherically symmetric and asymptotically flat BHs in the eikonal (large multipole number) regime are not determined by the parameters of the circular null geodesics, instead they are determined by the parameters of the circular photon orbit, since in the NED light ray does not follow the null geodesics~\cite{Novello:PRD:2000,Novello:PRD:2001,ObukhovPRD:66,Breto:PRD:2016,deOlivieraCQG:26}. Here, we approve that statement by study of the \textit{polar} EM perturbations. Below we briefly present the procedure.

In the large multipole numbers limit, one can write the effective potential (\ref{potential-el1}) in the following form:
\bear
V = \ell^2 \left[\frac{f}{r^2} \left( 1 -
\frac{ 4 \bar{\L}_{\bar{F}\bar{F}} \varphi'^2}{ \bar{\L}_{\bar{F}} }\right)^{- 1}
+ O \left(\frac{ 1 }{ \ell } \right) \right]\ ,\label{potential-eik}
\ear
The effective metric is constructed as
\bear
g_{eff}^{\mu \nu} = \bar{\L}_{\bar{F}} g^{\mu \nu}-4 \bar{\L}_{\bar{F}\bar{F}} g_{\alpha \beta} \bar{F}^{\mu \beta}\bar{F}^{\alpha \nu}\ .\label{effective-metric}
\ear
From (\ref{effective-metric}), one can construct an infinite number of the effective metrics with the conformal factors, such as
\bear\label{conformal1}
g^{eff}_{\mu\nu}=\Omega_1^2 diag\left\{-g,\frac{1}{h},r^2,r^2\sin^2\theta\right\}\ ,
\ear
with
\bear
g=f\left(1-\frac{4\bar{\L}_{\bar{F}\bar{F}} \varphi'^2}{\bar{\L}_{\bar{F}}}\right)^{-1},\quad h=f\left(1-\frac{4\bar{\L}_{\bar{F}\bar{F}} \varphi'^2}{\bar{\L}_{\bar{F}}}\right)\ ,\nonumber
\ear
or
\bear\label{conformal2}
g^{eff}_{\mu\nu}=\Omega_2^2 diag\left\{-f,\frac{1}{f},\tilde{r}^2,\tilde{r}^2\sin^2\theta\right\}\ ,
\ear
with
\bear
\tilde{r}=r\left(1-\frac{4\bar{\L}_{\bar{F}\bar{F}} \varphi'^2}{\bar{\L}_{\bar{F}}}\right)^{1/2}\ ,\nonumber
\ear
where the conformal factors $\Omega_1^2=1/\bar{\L}_{\bar{F}}$ and $\Omega_2^2=1/\left(\bar{\L}_{\bar{F}}-4\bar{\L}_{\bar{F}\bar{F}} \varphi'^2\right)$. It is well known that the conformal factor $\Omega^2$ can be ignored in the EM perturbations~\cite{ToshmatovPRD:96} and null
geodesics~\cite{BambiJCAP:5}, since it plays no role in these phenomena. In both (\ref{conformal1}) and (\ref{conformal2}) spacetime metrics, the effective potentials for the massless particles can be written as
\bear
V=g\frac{\ell^2}{r^2}\ , \quad or \quad V=f\frac{\ell^2}{\tilde{r}^2}\ ,
\ear
respectively, which are identical with (\ref{potential-eik}). Here, one can find the parameters of the circular massless particle orbit as
\bear
&&\Omega_c=\sqrt{\frac{g_c}{r_c^2}}\ ,\label{omega1}\\
&&\lambda=\sqrt{\frac{h_c}{2r_c^2}\left(2g_c-r_c^2g_c''\right)}\
,\label{lambda1}
\ear
or
\bear
&&\Omega_c=\sqrt{\frac{f_c}{\tilde{r}_c^2}}\ ,\label{omega2}\\
&&\lambda=\sqrt{\frac{f_c}{2\tilde{r}_c^2}\left(2f_c-\tilde{r}_c^2f_c''\right)}\
,\label{lambda2}
\ear
where the radius of the unstable photon orbit $r_c$ is determined by the equation $r_cg_c'-2g_c=0$ or $\tilde{r}_cf_c'-2f_c=0$.

\section{EM QNMs of the BHs in GR coupled to the NED}\label{sec-application}

In this section we apply the above demonstrated formalism for the new type of BH solutions in GR coupled to the NED derived in~\cite{TSA:PRDnew,TSSA:PRD:2018} reinterpreting the physical parameters of the spacetime presented in~\cite{Fan:PRD:2016}. Thus, the mass function of the metric function~(\ref{metric-function}) is given as
\bear\label{mass-function-new}
m(r)=M-\frac{q^3}{\alpha}\left[1 -\frac{r^{\mu}}{(r+q)^{\mu}}\right]\ ,
\ear
Where $\mu\geq3$ provides the spacetime to be regular only if the condition $M=M_{em}\equiv q^3/\alpha$ is satisfied. Therefore, the metric function of the new type of regular BH solutions are given as
\bear\label{metric-function-new}
f(r)=1-\frac{2Mr^{\mu-1}}{(r+q)^{\mu}}\ .
\ear
\begin{figure}[t]
\begin{center}
\includegraphics[width=0.9\linewidth]{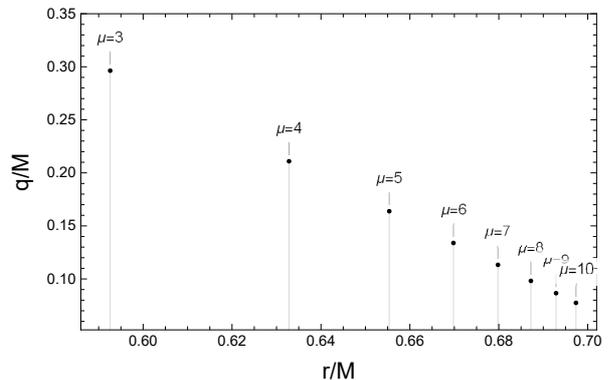}
\end{center}
\caption{\label{fig3} Boundaries of the Maxwellian BHs and no-horizon spacetimes in the parametric space. The black points correspond to the extremal BHs ($r_{ext}/M$, $q_{ext}/M$).}
\end{figure}
This solution represents the BH with two (inner and outer) horizons, extremal BH with only one degenerate horizon, and no-horizon spacetimes. These all cases depend on the values of parameters $q$ and $\mu$. In Fig.~\ref{fig3} these cases are presented in the parametric space. There shaded regions represent the possible values of the charge parameter for the spacetime to represent BHs.

Let us turn our attention to the EM perturbations of these solutions. Here one should note that the EM perturbations of the solution~(\ref{metric-function-new}) strongly depend on type of the charge of BHs, i.e., despite the geometry of the magnetically and electrically charged BHs~(\ref{metric-function-new}) are the same, their Lagrangian densities and consequently, their responses to the EM perturbations are significantly different as their effective potentials~(\ref{pot-mag1}) and~(\ref{potential-el1}) are different as well.

\begin{figure*}[t]
\begin{center}
\includegraphics[width=0.45\linewidth]{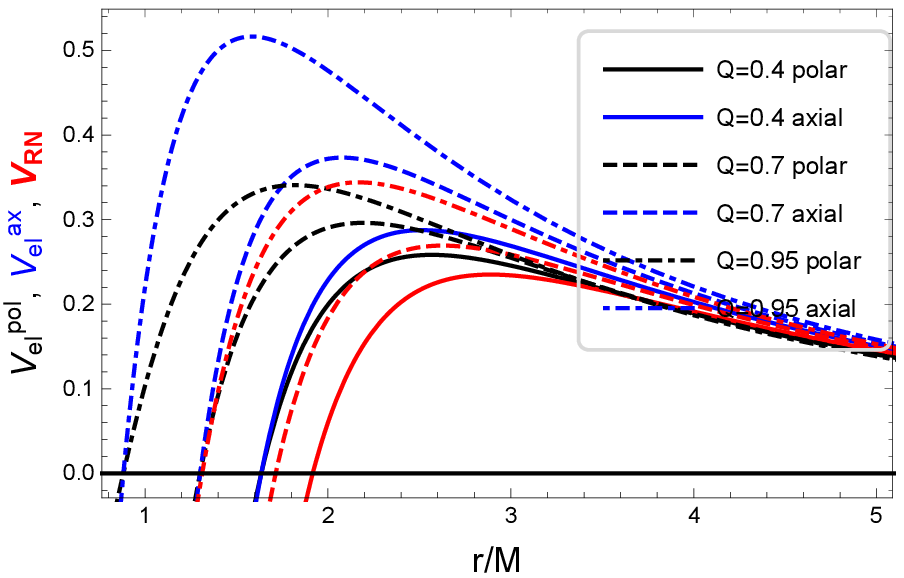}
\includegraphics[width=0.45\linewidth]{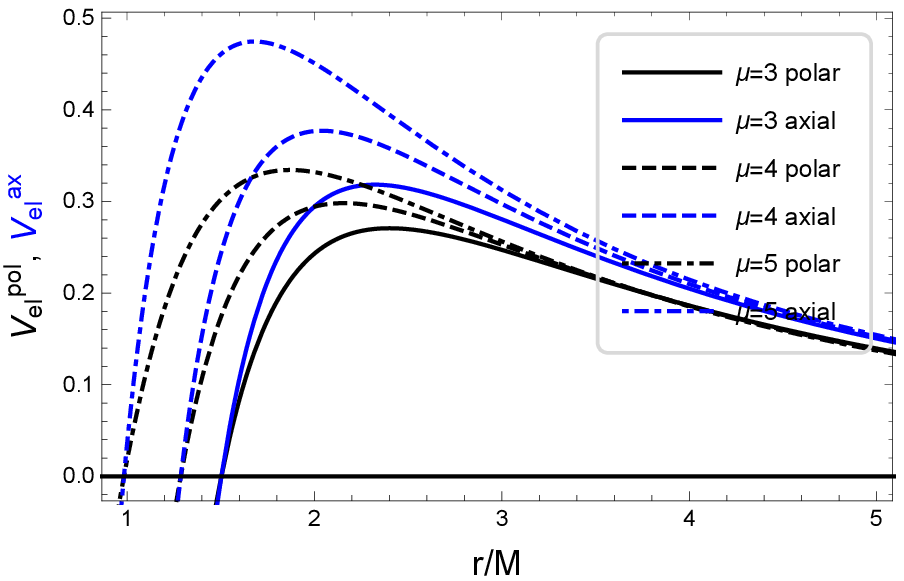}
\end{center}
\caption{\label{fig1} Effective potentials of the polar (black curves, given by the potential~(\ref{potential-el1})) and axial (blue curves, given by the potential~(34) of Ref.~\cite{TSSA:PRD:2018}) EM perturbations of the electrically charged regular BHs in comparison with the one of the RN BH (red curves) for the different values of the normalized charge parameter $Q$ (left panel) and $\mu$ (right panel).}
\end{figure*}

If we consider the BH solution~(\ref{line element}) with the mass function~(\ref{mass-function-new}) is electrically charged, then, the Lagrangian density of it is given by the function
\bear
\bar{\L}=\frac{2\mu}{\alpha}x^4(1+x)^{-\mu-2}((\mu-1)x-2)\ ,\quad x\equiv\frac{q}{r}\ ,
\ear
The electric potential is given as
\bear
\varphi=\frac{\mu}{\sqrt{2\alpha}}x(1+x)^{-\mu-1}(1-(\mu-3)x)\ ,
\ear
Then, the EM field strength takes the following form:
\bear
\bar{F}=-\frac{\mu^2}{\alpha}x^4(1+x)^{-2\mu-4}((\mu-3)x-4)^2\ .
\ear
The effective potential of the polar~(\ref{potential-el1}) and axial ((34) in the paper~\cite{TSSA:PRD:2018}) EM perturbations of the BH with the mass function~(\ref{metric-function-new}) are plotted in Fig.~\ref{fig1}. We do not report the full expression of these potentials because of their cumbersome forms. Where $\bar{\L}_{\bar{F}}=\bar{\L}'/\bar{F}'$, $\bar{\L}_{\bar{F}\bar{F}}=\bar{\L}_{\bar{F}}'/\bar{F}'$.

Now let us calculate the QNMs of the polar and axial EM perturbations of the electrically charged regular BHs~(\ref{line element}) with the mass function~(\ref{mass-function-new}) and compare them with the standard RN BH in the LED by the sixth order WKB method. Since the possible values of the charge parameters of the regular BH~(\ref{line element}) with the mass function~(\ref{mass-function-new}) and the RN BH are different, in order to facilitate the comparison we turn into the dimensionless normalized charge parameters as $q\rightarrow Q q_{ext}$ with $Q\equiv q/q_{ext}$. After this transformation charge parameters of the both BHs accept the values in the same range as $Q\in[0,1]$. One should note that the polar and axial EM perturbations of the electrically and magnetically charged RN BHs are the same, i.e.., they are isospectral.
\begin{table*}
\begin{ruledtabular}
\begin{tabular}{ccccccc}
$\ell$ & $Q$ & polar ($\mu=3$) & axial ($\mu=3$)  & RN BH &  polar ($\mu=4$) & axial ($\mu=4$) \\
\hline
   & 0 & 0.2459 - i 0.0931 & 0.2459 - i 0.0931 & 0.2459 - i 0.0931 & 0.2459 - i 0.0931 & 0.2459 - i 0.0931 \\
1 & 0.4 & 0.2689 - i 0.0950 & 0.2879 - i 0.0970 & 0.2540 - i 0.0940 & 0.2687 - i 0.0950 & 0.2856 - i 0.0967 \\
 & 0.7 & 0.2921 - i 0.0946 & 0.3375 - i 0.0982 & 0.2752 - i 0.0952 & 0.2918 - i 0.0945 & 0.3319 - i 0.0977 \\
  & 0.95 & 0.3173 - i 0.0898 & 0.4105 - i 0.0902 & 0.3177 - i 0.0904 & 0.3171 - i 0.0895 & 0.3988 - i 0.0898 \\ \hline
   & 0 & 0.4571 - i 0.0951 & 0.4571 - i 0.0951 & 0.4571 - i 0.0951 & 0.4571 - i 0.0951 & 0.4571 - i 0.0951 \\
2 & 0.4 & 0.4952 - i 0.0969 & 0.5269 - i 0.0987 & 0.4707 - i 0.0959 & 0.4949 - i 0.0968 & 0.5230 - i 0.0985 \\
 & 0.7 & 0.5332 - i 0.0966 & 0.6088 - i 0.0997 & 0.5056 - i 0.0970 & 0.5328 - i 0.0965 & 0.5994 - i 0.0992 \\
  & 0.95 & 0.5755 - i 0.0923 & 0.7292 - i 0.0916 & 0.5751 - i 0.0927 & 0.5751 - i 0.0920 & 0.7097 - i 0.0913 \\ \hline
   & 0 & 0.6567 - i 0.0956 & 0.6567 - i 0.0956 & 0.6567 - i 0.0956 & 0.6567 - i 0.0956 & 0.6567 - i 0.0956 \\
3 & 0.4 & 0.7101 - i 0.0974 & 0.7546 - i 0.0992 & 0.6758 - i 0.0965 & 0.7097 - i 0.0974 & 0.7491 - i 0.0990 \\
 & 0.7 & 0.7632 - i 0.0971 & 0.8691 - i 0.1001 & 0.7245 - i 0.0976 & 0.7625 - i 0.0970 & 0.8560 - i 0.0997 \\
  & 0.95 & 0.8224 - i 0.0930 & 1.0374 - i 0.0919 & 0.8215 - i 0.0933 & 0.8219 - i 0.0927 & 1.0100 - i 0.0917 \\
\end{tabular}
\end{ruledtabular}
\caption{\label{tab1} Fundamental ($n=0$) QNMs of the polar and axial EM perturbations of the electrically charged BH~(\ref{line element}) with the mass function~(\ref{mass-function-new}) in the NED in comparison with the ones of the RN BHs in the LED.}
\end{table*}

In Table~\ref{tab1} and Fig.~\ref{fig-qnm} some fundamental QNMs of the axial and polar EM perturbations of the electrically charged BH~(\ref{line element}) with the mass function~(\ref{mass-function-new}) in the NED in comparison with the ones of the RN BHs in the LED are presented for the several values of the electric charge parameter $q$ and nonlinearity degree $\mu$. In Fig.~\ref{fig-qnm} the junctions of the curves correspond to value of the QNMs of the Schwarzschild BH.

\begin{figure*}[t]
\begin{center}
\includegraphics[width=0.45\linewidth]{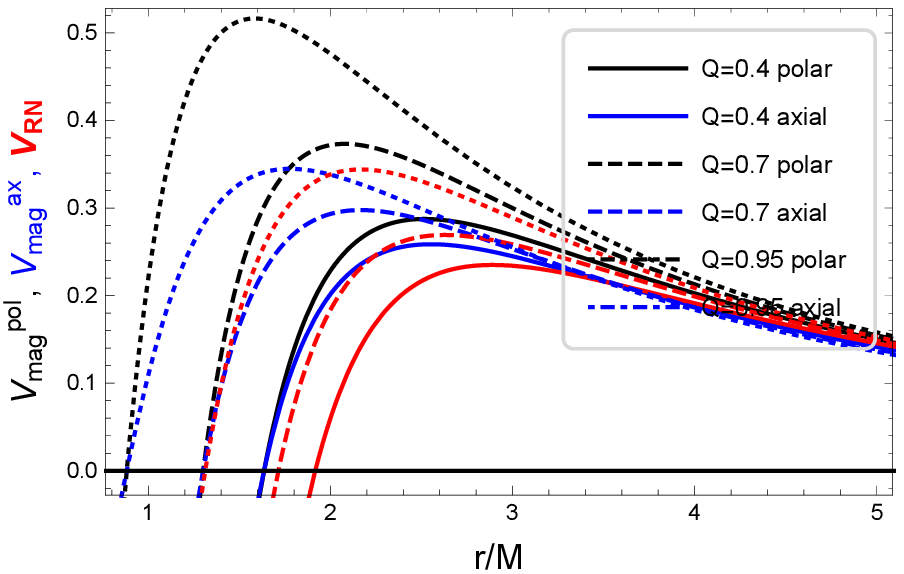}
\includegraphics[width=0.45\linewidth]{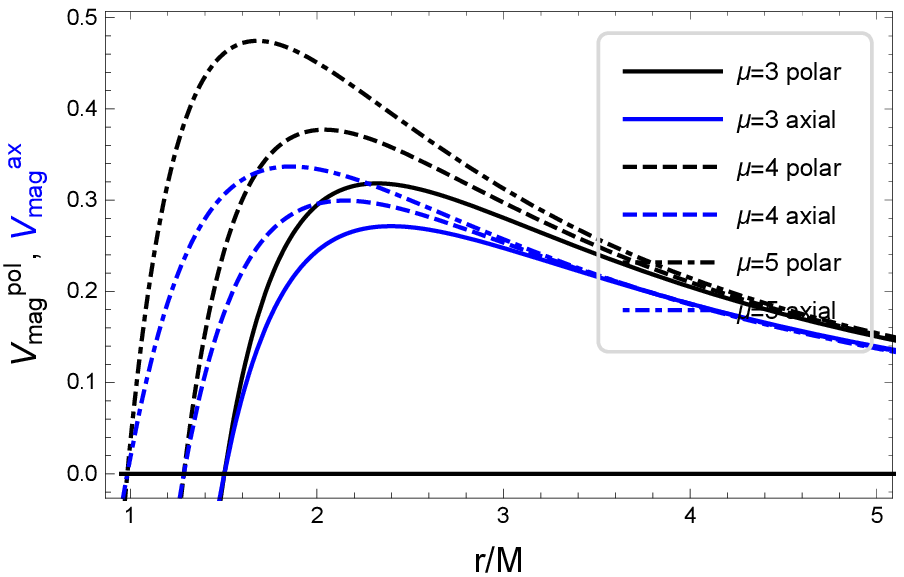}
\end{center}
\caption{\label{fig2} The same as Fig.~\ref{fig1} but for the magnetically charged regular Maxwellian BHs. Where black curves correspond to the potential~(\ref{pot-mag1}), while the blue curves correspond to the potential (43) of Ref.~\cite{TSSA:PRD:2018}.}
\end{figure*}
\begin{table*}
\begin{ruledtabular}
\begin{tabular}{ccccccc}
$\ell$ & $Q$ & polar ($\mu=3$) & axial ($\mu=3$)  & RN BH &  polar ($\mu=4$) & axial ($\mu=4$) \\
\hline
   & 0 & 0.2459 - i 0.0931 & 0.2459 - i 0.0931 & 0.2459 - i 0.0931 & 0.2459 - i 0.0931 & 0.2459 - i 0.0931 \\
1 & 0.4 & 0.2876 - i 0.0970 & 0.2691 - i 0.0950 & 0.2540 - i 0.0940 & 0.2853 - i 0.0968 & 0.2689 - i 0.0950 \\
 & 0.7 & 0.3366 - i 0.0984 & 0.2928 - i 0.0945 & 0.2752 - i 0.0952 & 0.3311 - i 0.0978 & 0.2924 - i 0.0943 \\
  & 0.95 & 0.4085 - i 0.0907 & 0.3187 - i 0.0887 & 0.3177 - i 0.0904 & 0.3970 - i 0.0900 & 0.3180 - i 0.0884 \\ \hline
   & 0 & 0.4571 - i 0.0951 & 0.4571 - i 0.0951 & 0.4571 - i 0.0951 & 0.4571 - i 0.0951 & 0.4571 - i 0.0951 \\
2 & 0.4 & 0.5268 - i 0.0987 & 0.4953 - i 0.0969 & 0.4707 - i 0.0959 & 0.5230 - i 0.0985 & 0.4950 - i 0.0968 \\
 & 0.7 & 0.6084 - i 0.0997 & 0.5336 - i 0.0964 & 0.5056 - i 0.0970 & 0.5991 - i 0.0993 & 0.5330 - i 0.0964 \\
  & 0.95 & 0.7281 - i 0.0917 & 0.5760 - i 0.0918 & 0.5751 - i 0.0927 & 0.7086 - i 0.0914 & 0.5753 - i 0.0916 \\ \hline
   & 0 & 0.6567 - i 0.0956 & 0.6567 - i 0.0956 & 0.6567 - i 0.0956 & 0.6567 - i 0.0956 & 0.6567 - i 0.0956 \\
3 & 0.4 & 0.7545 - i 0.0992 & 0.7102 - i 0.0974 & 0.6758 - i 0.0965 & 0.7491 - i 0.0990 & 0.7097 - i 0.0974 \\
 & 0.7 & 0.8688 - i 0.1001 & 0.7634 - i 0.0970 & 0.7245 - i 0.0976 & 0.8557 - i 0.0997 & 0.7626 - i 0.0969 \\
  & 0.95 & 1.0366 - i 0.0920 & 0.8227 - i 0.0927 & 0.8215 - i 0.0933 & 1.0093 - i 0.0917 & 0.8219 - i 0.0925 \\
\end{tabular}
\end{ruledtabular}
\caption{\label{tab2} The same as Table~\ref{tab1} but for the magnetically charged Maxwellian BHs.}
\end{table*}

\begin{figure*}[t]
\begin{center}
\includegraphics[width=0.45\linewidth]{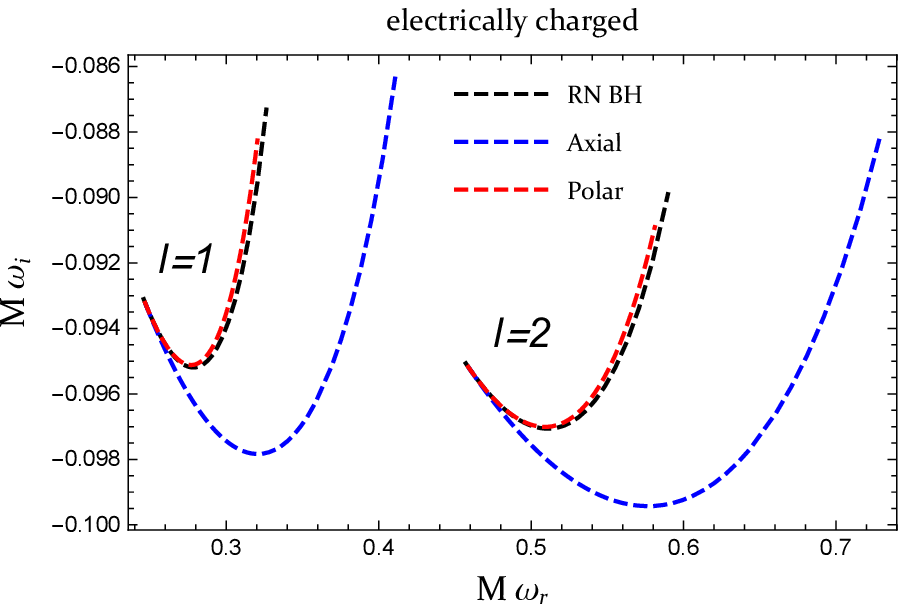}
\includegraphics[width=0.45\linewidth]{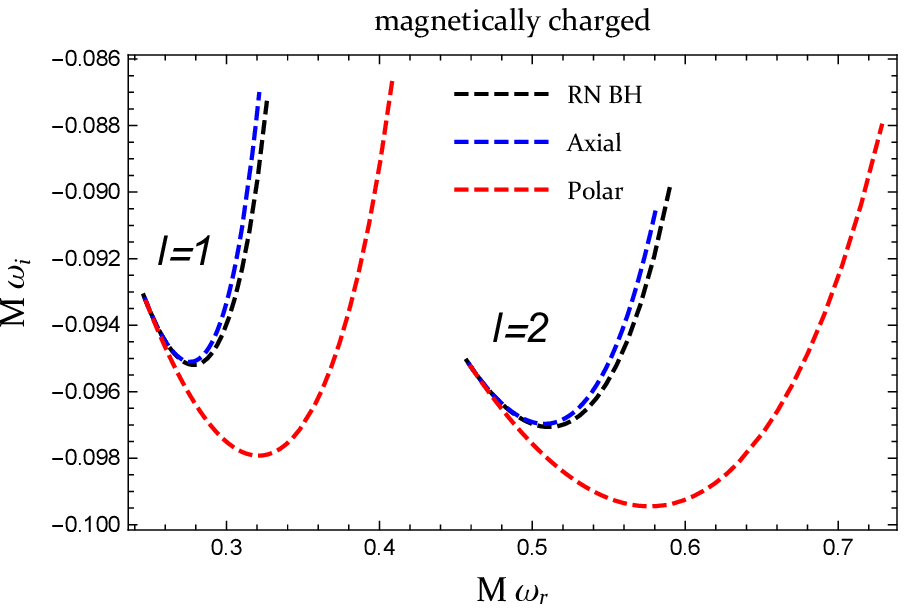}
\end{center}
\caption{\label{fig-qnm} $\ell=1$ and $\ell=2$ fundamental QNMs of the EM perturbations of the electrically and magnetically charged (Maxwellian) regular BH in the NED and RN BH in the LED for the whole range of the charge parameter $Q\in[0,1]$. Where junctions of the curves correspond to the ones of the Schwarzschild BH ($Q=0$).}
\end{figure*}

If the magnetically charged BH solutions in the GR coupled to the NED~(\ref{line element}) with the mass function~(\ref{mass-function-new}) are considered, we will not present the Lagrangian densities of that solutions since in our previous paper~\cite{TSSA:PRD:2018} they have been demonstrated. Since in this case the Lagrangian density tends to the Maxwell one in weak field regime, we called it as Maxwellian BHs. Moreover, again we do not report the full expressions of the effective potentials of the polar and axial EM perturbations due to their cumbersome forms, instead we show their forms in Fig.~\ref{fig2}. In Table~\ref{tab2} and Fig.~\ref{fig-qnm} the QNMs of the polar and axial EM perturbations of the magnetically charged regular BHs in the NED and RN BH in the LED have been presented.

One can see from Tables~\ref{tab1},~\ref{tab2} and Fig.~\ref{fig-qnm} that with increasing the value of the charge parameters the frequencies of the real oscillations of the EM perturbations of the BHs in the NED and LED increase, while the damping rates of these oscillations increase up to $Q\approx0.7$, then they decrease dramatically. One of the momentous results of this paper is that if the geometry (spacetime) of the electrically and magnetically charged BHs in the NED are the same, the axial EM perturbations of the magnetically charged and polar EM perturbations of the electrically charged BHs are isospectral, while axial EM perturbations of the electrically charged and polar EM perturbations of the magnetically charged BHs are isospectral as
\bear\label{phenomeno}
\omega_{mag}^{ax}\approx\omega_{el}^{pol}, \qquad
\omega_{mag}^{pol}\approx\omega_{el}^{ax},
\ear
i.e., without specifying the type of the EM perturbations, one cannot recognize if the BH in the GR coupled to the NED is magnetically or electrically charged, and vice versa, from the QNM spectra.

Moreover, an increase in the value of parameter $\mu$ decreases both real and imaginary parts of the QNMs. Thus, one can conclude that the EM perturbations of highly charged BHs or BHs in the NED with big $\mu$ parameters live longer. Furthermore, the EM perturbations of the BH in the LED (RN BH) live longer than in the NED for the small and intermediate values of the charge. However, for the large values of the charge parameter the EM perturbations of the BH in the NED  live longer.
\begin{figure}[t]
\begin{center}
\includegraphics[width=0.9\linewidth]{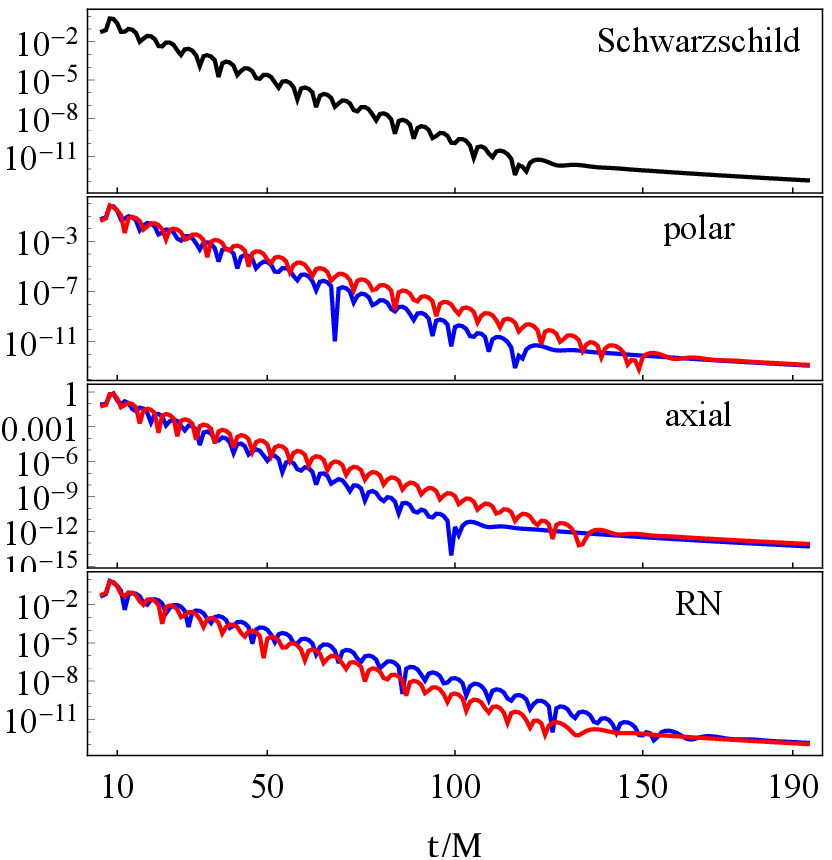}
\end{center}
\caption{\label{fig5} Semilog plot of the temporal evolution of $\ell=2$ axial and polar EM perturbations of the electrically charged BH~(\ref{line element}) with the mass function~(\ref{mass-function-new}) in the NED in comparison with the RN BH in the LED and the Schwarzschild BH. Where blue and red curves correspond to $Q=0.7$ and $Q=0.95$, respectively. }
\end{figure}
In Fig.~\ref{fig5} time domain profile of the EM perturbations of the electrically charged BH~(\ref{line element}) with the mass function~(\ref{mass-function-new}) for the different values of the charge parameter have been presented in comparison with the ones of the Schwarzschild and RN BHs. Since from Fig.~\ref{fig-qnm} and Eq.~(\ref{phenomeno}) it can be realized easily that the time domain profile of the EM perturbations of the magnetically charged BH~(\ref{line element}) with the mass function~(\ref{mass-function-new}) for the different values of the charge parameter is the same as Fig.~\ref{fig5} only if axial and polar epilogs are interchanged.

\section{Conclusion}\label{sec-conclude}

The present paper represents a continuation of our recent paper~\cite{TSSA:PRD:2018} devoted to the study of the \emph{axial} EM perturbations of the BHs in GR coupled to the NED, considered for both electrically and magnetically charged BHs under assumption that the EM perturbations do not alter the spacetime geometry. The detailed analysis of the \emph{polar} EM perturbations performed in the present study have demonstrated that the polar EM perturbations of the NED BHs give different effective potentials and consequently, different results for the QNMs, as compared to those related to the standard RN BHs in the LED. It is well known that both the \emph{axial} and \emph{polar} EM perturbations of the electrically and magnetically charged BHs in the LED (RN) are isospectral, i.e., they have the same effective potentials and QNMs. However, in the case of the BHs in the NED, electrically and magnetically charged BHs have different potentials and different QNM spectra.

Moreover, we have shown in the detailed study of the QNMs properties in the eikonal (large multipole numbers) approximation that the EM perturbations can play a powerful tool to confirm that the light ray does not follow the null geodesics of the spacetime in the NED. To be more precise, in the paper~\cite{CardosoPRD:79} it was formulated that in the eikonal regime QNMs are determined by the parameters of the unstable circular null geodesics. In this paper, analysis of the \emph{polar} EM perturbations (as \emph{axial} EM perturbations in~\cite{TSSA:PRD:2018}) have shown that the QNMs of the BHs in the NED are determined by the unstable circular photon orbits determined by the effective geometry, as in the case of the axial perturbations.

As a special case we have studied the polar EM perturbations of the electrically and magnetically charged new BH solutions in GR coupled to the NED~\cite{TSA:PRDnew} in comparison with the ones of the RN BH in the LED and Schwarzschild BH. Moreover, we have compared the obtained polar EM perturbations with the known axial EM perturbations~\cite{TSSA:PRD:2018}. The detailed analysis of the QNM spectra of the axial and polar EM perturbations of the electrically and magnetically charged BHs provide justification for fundamental statement that the magnetically and electrically charged BH spacetimes in the GR coupled to the NED are dual to each other, i.e. axial EM perturbations of magnetically (electrically) charged BH and polar EM perturbations of the electrically (magnetically) charged BH are isospectral, i.e. $\omega_{mag}^{ax}\approx\omega_{el}^{pol}$ ($\omega_{mag}^{pol}\approx\omega_{el}^{ax}$).

\section*{Acknowledgments}

B.T. is grateful to Bobur Turimov for his help on plotting figures. B.T. and Z.S. would like to acknowledge the institutional support of the Faculty of Philosophy and Science of the Silesian University in Opava, the internal student grant of the Silesian University Grant No.~SGS/14/2016 and the Albert Einstein Centre for Gravitation and Astrophysics under the Czech Science Foundation Grant No.~14-37086G. The work was supported by Nazarbayev University Faculty Development Competitive Research Grant: ``Quantum gravity from outer space and the search for new extreme astrophysical phenomena", Grant No.~090118FD5348. The researches of B.A. and B.T. are partially supported by Grants No.~VA-FA-F-2-008 and No.~YFA-Ftech-2018-8 of the Uzbekistan Ministry for Innovation Development, by the Abdus Salam International Centre for Theoretical Physics through Grant No.~OEA-NT-01 and by an Erasmus+ exchange grant between Silesian University in Opava and National University of Uzbekistan.

\label{lastpage}

\bibliography{Toshmatov_references}

\end{document}